\documentclass[traditabstract]{aa}
\usepackage{graphicx}
\usepackage{txfonts}
\usepackage[utf8]{inputenc}
\usepackage{natbib}
\usepackage[dvipsnames]{xcolor}
\usepackage{lscape}
\usepackage{amsmath}
\usepackage{textcomp}

\newcommand{\integral}{\textit{INTEGRAL}\xspace}
\newcommand{\eqpair}{\textsc{eqpair}\xspace}
 
\begin{document}

   \title{\integral discovery of a high-energy tail in the microquasar Cygnus X-3}
   \subtitle{}

   \author{F. Cangemi\inst{1},
   J. Rodriguez\inst{1},
   V. Grinberg\inst{2},
   R. Belmont\inst{1},
   P. Laurent\inst{1},
   \and J. Wilms\inst{3}}

\offprints{floriane.cangemi@cea.fr}
\authorrunning{Cangemi et al.}
\titlerunning{Cyg X-3 high-energy tails}

   \institute{Lab AIM, CEA/CNRS/Universit\'e Paris-Saclay, Universit\'e de Paris, F-91191 Gif-sur-Yvette, France\\
             \email{floriane.cangemi@cea.fr, fcangemi@lpnhe.in2p3.fr}
             \and Institut f\"ur Astronomie und Astrophysik, Universität T\"ubingen, Sand 1, 72076 T\"ubingen, Germany
             \and Dr.~Karl Remeis-Sternwarte and Erlangen Centre for Astroparticle Physics, Friedrich-Alexander Universit\"at Erlangen-N\"urnberg, Sternwartstr.~7, 96049 Bamberg, Germany
        }
   \date{Accepted}

  \abstract
  {The X-ray spectra of X-ray binaries are dominated by emission of  either soft or hard X-rays which defines their soft and hard spectral states. While the generic picture is relatively well understood, little is known about the interplay of the various media at work, or about the reasons why some sources do not follow common behavior. Cygnus X-3 is amongst the list of X-ray binaries that show quite complex behavior, with various distinct spectral states not seen in other sources. These states have been characterized in many studies. Because of its softness and intrinsic low flux above typically 50\,keV, very little is known about the hard X/soft gamma-ray (100--1000\,keV) emission in Cygnus X-3.}
  {Using the whole INTEGRAL data base, we aim to explore the 3--1000\,keV spectra of Cygnus X-3. This allows to probe this region with the highest sensitivity ever, and search for the potential signature of a high-energy non-thermal component as sometimes seen in other sources.}
  {Our work is based on state classification carried out in previous studies with data from the \textit{Rossi X-Ray Timing Explorer}. We extend this classification to the whole \integral\ data set in order to perform a long-term state-resolved spectral analysis. Six stacked spectra were obtained using 16 years of data from JEM-X (3--25\,keV), ISGRI (25--300\,keV), and SPI (20--400\,keV).}
  {We extract stacked images in three different energy bands, and detect the source up to 200\,keV. In the hardest states, our purely phenomenological approach clearly reveals the presence of an additonnal component $> 50$\,keV in addition to the component usually interpreted as thermal Comptonization. We apply a more physical model of hybrid thermal/nonthermal corona (\textsc{eqpair}) to characterize this nonthermal component and compare our results with those of previous studies and analyses. Our modeling indicates a more efficient acceleration of electrons in states where major ejections are observed. We also evaluate and find a dependence of the photon index  of the power law as a function of the strong orbital modulation of the source in the Flaring InterMediate (FIM) state. This dependence could be due to a higher absorption when Cygnus X-3 is behind its companion. However, the uncertainties on the density column prevent us from drawing any firm conclusions.}
  {}
   \keywords{Accretion, accretion disks --- Physical data and processes --- X-rays: binaries --- Radiation mechanisms: non-thermal --- Stars: individual (Cygnus X-3)}

   \maketitle

\section{Introduction}
\label{intro}

During their outburst, X-Ray binaries (XRBs) can pass through different accretion states associated with intrinsic emitting properties that are drastically different. Transient XRBs spend the majority of their life in a quiescent state, before entering into a period of outburst in the so-called hard state. Here, the spectrum is dominated by emission in the hard ($\sim$10--100\,keV) X-rays: the commonly accepted interpretation is that of an inverse Compton-scattering of soft photons emitted by a cold ($\leq0.1$\,keV) accretion disk by hot electrons (50--100\,keV) forming a hot \textquotedblleft corona\textquotedblright\ \citep{Haard1991}. This state is also associated with a compact jet detected in the radio domain \citep[e.g.,][]{Fender2001, Stirling2001, Fuchs2003, Corbel2013}. On the contrary, after the hard state, XRBs are found in a soft state with a spectrum dominated by thermal emission in the soft ($\sim$1\,keV) X-rays. The disk is thought to be closer to the compact object, the jet is quenched \citep[e.g.,][]{Fender1999b, Corbel2000}, and the Comptonized emission is much weaker, possibly indicating the disappearance of the corona itself \citep{Rodriguez2003, Rodriguez2008II}. The transition from the hard to the soft state is made through the so-called intermediate states \citep{Belloni2005} with discrete and sometimes superluminal radio ejections marking the hard--soft frontier \citep[][]{Tingay1995}.

Beyond a few hundred keV the picture is much more blurred. For decades, the weak flux of the sources and the lack of sensitive instruments have  prevented efforts to detect them, or  to probe the eventual connections of the $>100$\,keV emission with the X-ray states. Observations with the \textit{Compton Gamma-Ray Observatory} \citep[\textit{CGRO},][]{Grove1998, McConnell2000, Gierlinski2003} and the \textit{INTernational Gamma-Ray Astrophysics Laboratory} \citep[\integral,][]{Joinet2007, Laurent2011, Tarana2011, Jourdain2012, Jourdain2014, Rodriguez2015} have nevertheless shown that hard-X-ray excesses beyond 100\,keV, so called high-energy tails, are common in black hole XRBs. Even for the prototypical black hole XRB, Cygnus X-1 (Cyg X-1), for which a high-energy tail was confirmed very early in the \integral\,lifetime
\citep{Bouchet2003, Cadolle2006}, the origin of the tail is not
yet well understood; it could either come from synchrotron emission from the basis of the jets \citep{Laurent2011, Rodriguez2015}, or from hybrid thermal/nonthermal electron distribution in the corona \citep[e.g.,][]{DelSanto2013, Romero2014}, or both depending on the state (Cangemi et al., submitted to A\&A.). By studying the presence and behavior of hard tails in other similar sources, we hope to understand the commonalities and define the origin of these features. In this paper, we investigate the case of the very bright source Cygnus X-3 (Cyg X-3) and study the potential presence of a hitherto undetected high-energy tail.

Cyg X-3 is one of the first discovered XRBs \citep{Giacconi1967}. Its nature  still remains a mystery, because for this compact object in particular, it is extremely difficult to obtain the mass function of the system \citep[e.g.,][]{Hanson2000, Vilhu2009}. However, its global behavior, the various spectra, and their properties seem to indicate a black hole rather than a neutron star \citep[e.g.,][H09]{Szostek2008, Zdziarski2013, Koljonen2017, Hjalmarsdotter2009}. In this system the compact object is extremely close to its Wolf-Rayet companion \citep{VanKerkwijk1992, Koljonen2017} rendering the system peculiar in many ways when compared to other XRBs with low companion mass, such as for example GX 339--4 or even the high-mass system Cyg X-1. It is situated at a distance of $7.4\pm 1.1$\,kpc \citep{McCollough2016} with an orbital period of 4.8\,h \citep{Parsignault1972}.

Cyg X-3 is a microquasar owing to the presence of strong radio flares, and is the brightest radio source of this kind \citep{McCollough1999}. These radio properties agree with those expected from compact jets \citep[e.g.,][]{Schalinski1995, Molnar1988, Mioduszewski2001, Miller-jones2004, Tudose2007, Egron2017} but Cyg X-3 also shows discrete ejections during flares. These jets are very variable, as indicated by the fast variations in radio \citep{Tudose2007}, and for this reason the source has its own states defined by their radio flux: \textit{quiescent}, \textit{minor flares}, \textit{major flares} which occur after a period of \textit{quenched} emission \citep{Waltman1996}. At higher energies, Cyg X-3 has been detected by Fermi in the $\gamma$-ray range and its flux is positively correlated with the radio flux while showing variations correlated with the orbital phase \citep{Fermi2009}. During flaring states, the $\gamma$-ray spectrum seems to be well modeled by Compton scattering of the soft photons from the companion by relativistic electrons from the jets \citep{Dubus2010, Cerutti2011, Zdziarski2018}.

Cyg X-3 shows a wider variety of states than the two canonical ones defined above. While the overall shape of its spectra is similar to those of other black hole XRBs, the value of the spectral parameters can be markedly different: the exponential cutoff is at a lower energy of $\sim$20\,keV in the hardest states, whereas the disk is very strong in the softest states. Cyg X-3
also shows a strong iron line and very strong absorption \citep{Szostek2008b}. This complexity and its correlation with the radio behavior led to the definition of five X-ray states when considering the spectral shapes and levels of fluxes \citep[][hereafter S08]{Szostek2008}, plus a `hypersoft' one when one uses the hardness intensity diagram \citep[HID,][hereafter K10]{Koljonen2010}. The latter state is modeled with a pure 1.5 keV black-body spectrum and a $\Gamma \sim2.5$\footnote{where $\Gamma$ is the power-law photon index such that the photon number density is $dN(E)\propto E^{-\Gamma} dE$.} power law to represent the hard X-ray emission. Here we give a brief description of these states, from the hardest to the softest (see K10 Sect. 4.2 for more details):

\textit{Quiescent} state. This state is characterized by a high flux in the hard X-rays and a low flux in the soft X-rays. The radio flux is about 60-200 mJy (K10) and anticorrelates with hard X-rays (20--100\,keV) but correlates with soft X-rays (3--5\,keV, S08). The spectrum appears to be well fitted by Comptonization models with an exponential cutoff around 20\,keV and a strong iron line at 6.4\,keV (e.g., K10).

\textit{Transition} state. In this state, the radio and soft X-ray fluxes start to increase. The source starts to move to the left part (soft) of the HID. However, the hard X-rays have a spectral shape that is quite similar to that of the quiescent state. This state and the quiescent state would correspond to the hard state in a standard black hole XRB, such as GX 339--4 \citep{Belloni2005}. The quiescent and transition states correspond to the radio \textit{quiescent} state, with a typical radio flux of $\sim 130$\,mJy \citep[e.g.,][K10]{Szostek2008}.

\textit{Flaring hard X-ray (FHXR)} state. The shape of the spectrum starts to soften significantly in this state. Minor flaring is observed in the radio. This state would correspond to the intermediate state in a standard black hole XRB, and corresponds to the radio \textit{minor flaring} state \citep[with a mean radio flux of $\sim 250$\,mJy, e.g.,][K10]{Szostek2008}

\textit{Flaring intermediate (FIM)} state. Major flares are observed, the spectrum is softer than in the FHXR and the presence of the disk starts to clearly appear in the spectrum. This state would correspond to a soft/intermediate state in a standard black hole XRB.

\textit{Flaring soft X-rays (FSXR)} state. FSXR and hypersoft states seem similar in terms of spectra. However, we observe a higher radio flux in the FSXR state whereas it is very low in the hypersoft state. These two states are separated by the \textquotedblleft\,jet-line\textquotedblright\,(K10), and unlike other black hole XRBs, a major flare occurs when the source goes from the hypersoft state to the FSXR state. 
The FSXR and FIM correspond to the radio \textit{major flaring} state \citep[300\,mJy to $\sim 10$\,Jy, e.g.,][K10]{Szostek2008}.

\textit{Hypersoft} state. The radio flux is almost quenched in this state \citep[$\sim 10$\,mJy, e.g.,][K10]{Szostek2008}. We see a strong presence of the disk in the spectrum while no emission above 80\,keV has been reported so far (H09, K10). The hypersoft state corresponds to the \textit{quenched} radio state.

Despite a huge number of observations both in soft X-rays (1--10\,keV) and hard (10--150\,keV) X-rays (notably with \textit{RXTE}), and contrary to many other bright XRBs \citep[e.g.,][Cangemi et al., in prep.]{Grove1998, Joinet2007, Rodriguez2008, Laurent2011, DelSanto2016}, only one detection between 100 and 200\,keV has been reported so far in Cyg X-3 (H09). Here we make use of \integral to probe the properties of this peculiar source over the full 3--1000\,keV range covered by the observatory. We use the spectral classification of K10 to separate the data into the six  spectral states defined therein to extract state-resolved stacked spectra. The description of the observations, and the data-reduction methods are reported in Sect. \ref{data_reduc}. Section \ref{classification} is dedicated to the state classification of the \integral data. We then present a phenomenological approach to the spectral fitting in Section \ref{phenom}, before considering more physical models in Section \ref{sec:eqpair}. The results are discussed in the last part in Sect. \ref{discussion}.

\section{\integral observations and data reduction}
\label{data_reduc}
\subsection{Data selection}
\label{Data_selection}

We consider all \integral individual pointings or science windows\footnote{One single \integral observation with a typical duration of $\sim$1500 ks.} (scws) of Cyg X-3 since the launch of \textit{INTEGRAL}  in 2002. We restrict our selection to scws where the source is in the field of view of the Joint European X-ray Monitors \citep[JEM-X,][]{Lund2003} in order to be able to use the soft X-ray classification with JEM-X data, that is, where the source is less than 5$^{\circ}$ off-axis. As JEM-X is composed of two units that have not always
been observing at the same time, this selection results in 1518 scws for JEM-X 1 and 185 scws for JEM-X 2. For scws with both JEM-X 1 and JEM-X 2 on, we only select the JEM-X 1 spectrum in order to avoid accumulating double scws.
In this paper, all the following scientific products were extracted on a scw basis, before being stacked in a state-dependent fashion.
We exclude \integral revolutions 1554, 1555, 1556, 1557, and 1558 to avoid potential artifacts caused by the extremely bright \citep[up to $\sim50$ Crabs, e.g.,][]{Rodrigue2015_V404} flares of V404 Cygni during its 2015 June outburst.

\subsection{\integral/JEM-X spectral extraction}
\label{jemx}

After the selection of the scws according to the criteria defined in Sect. \ref{Data_selection}, the data from JEM-X are reduced with version 11 of \textit{INTEGRAL} Off-line Scientific Analysis (OSA) software. We follow the standard steps described in the JEM-X user manual\footnote{\url{https://www.isdc.unige.ch/integral/download/osa/doc/11.0/osa_um_jemx/man.html}.}. Although, as mentioned above, most of the data are obtained with JEM-X unit 1, we nevertheless extract all data products from any of the units that were turned on during a given scw. Spectra are extracted in each scw where the source is automatically detected by the software at the image creation and fitting step. Spectra are computed over 32 spectral channels using the standard binning definition.

The individual spectra are then combined with the OSA \textit{spe\_pick} tool according to the classification scheme described hereafter in Sect. \ref{integral_class}.  In some cases, JEM-X background calibration lines seem to affect the spectra (C.-A. Oxborrow and J. Chenevez private communication). This effect is particularly obvious in bright, off-axis sources, and is thus amplified when dealing with large chunks of accumulated data. To avoid this problem, when particularly obvious, we omitted the JEM-X spectral channels at these energies. This is particularly evident in the FIM, FSXR, and hypersoft data where the source fluxes at low energies are the highest. The appropriate ancillary response files (arfs) are produced during the spectral extraction and combined with \textit{spe\_pick}, while the redistribution matrix file (rmf) is rebinned from the instrument characteristic standard rmf with \textit{j\_rebin\_rmf}. We add 3\,\% systematic error onto all spectral channels for each of the stacked spectra, as recommended in the JEM-X user manual. We determine the net count rates in the 3--6, 10--15, and 3--25\,keV ranges normalized to the on-axis value for each individual spectrum.

\begin{figure*}[h]
\centering
\includegraphics[width=\textwidth]{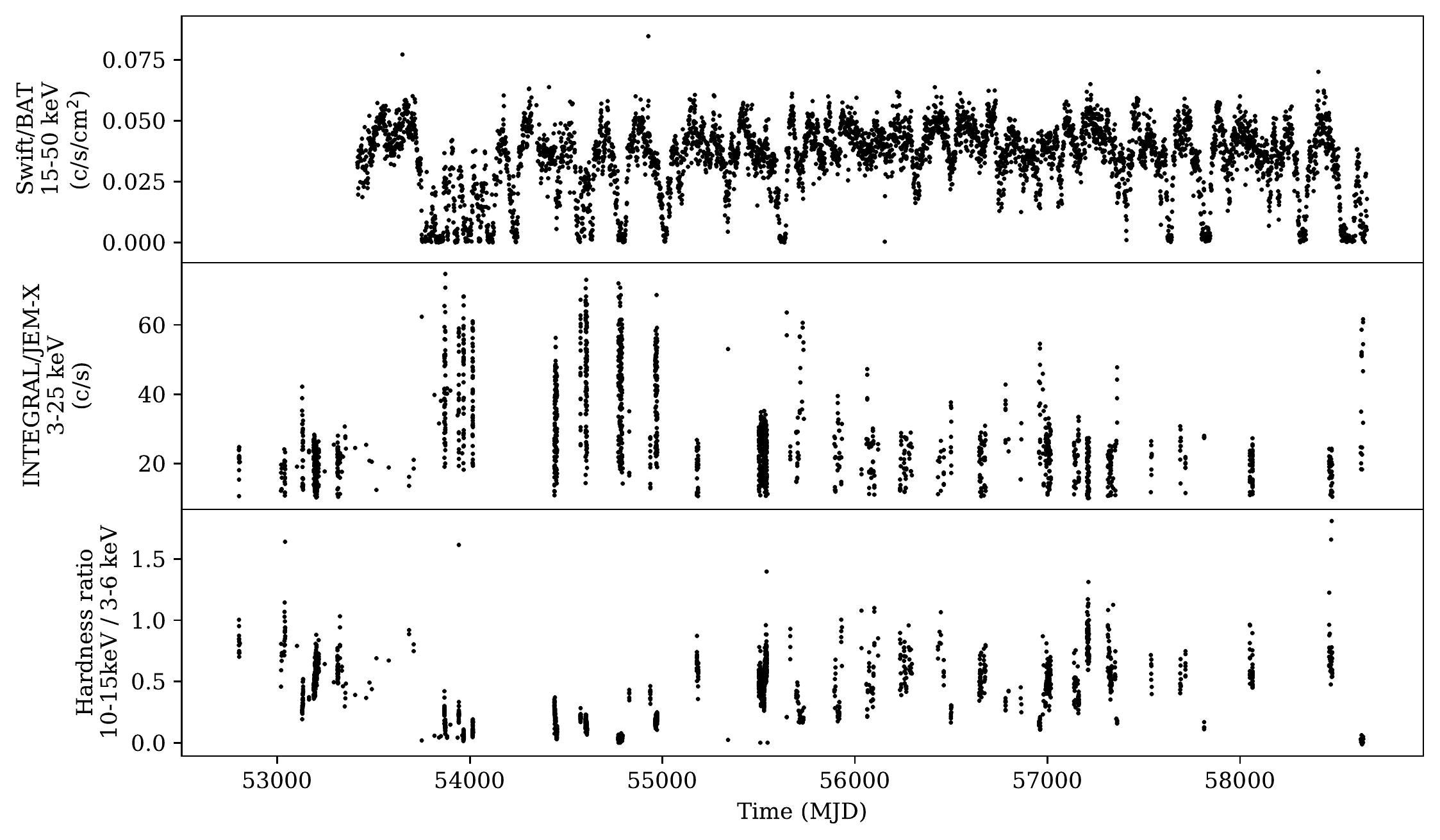}
\caption{Cyg X-3 light curves and hardness ratio. \textit{Upper panel:} \textit{Swift}/BAT daily light curve in its 15--50\,keV band. \textit{Middle Panel:} JEM-X scw basis light curve in its 3--25\, keV band. \textit{Lower panel:} JEM-X scw basis hardness ratio. The error bars are smaller than the size of the symbol.}
\label{lc_jemx}
\end{figure*}

Table \ref{nb_scws} indicates the number of scws for each state for both JEM-X 1 and 2. The upper panel of Fig.  \ref{lc_jemx} shows the daily \textit{swift}/BAT light curve, whereas the JEM-X 3-25\,keV band light curve and the hardness ratio are shown in the middle and lower panels, respectively. The hardness ratio shows the spectral variability of the source and we observe transition to the softest states observed by \integral around MJD 54000 when the JEM-X count rate reaches its maximum. Another transition to a very low hardness ratio is seen with \integral around MJD 58530 \citep{Atel2019}.

\subsection{\integral/IBIS/ISGRI spectral extraction}

To probe the behavior of  source in the hard X-rays we make use of data from the first detector layer of the Imager on Board the \integral Satellite (IBIS), the \integral Soft Gamma-ray Imager (ISGRI), which is sensitive between $\sim$20 and $\sim$600\,keV \citep{Lebrun2003}. As OSA version 11.0 is valid for ISGRI data taken since January 1, 2016 (MJD 57388), we divide our analysis in two parts\footnote{\url{https://www.isdc.unige.ch/integral/analysis#Software}}. The first part, which is analyzed with OSA 10.2 extends from MJD 52799 (\integral revolution 80) to MJD 57361 (rev 1618), while the second part is analyzed with OSA 11.0 and extends from MJD 57536 (rev 1684) to MJD 58639 (rev 2098).

Light curves and spectra are extracted following standard procedures\footnote{\url{https://www.isdc.unige.ch/integral/download/osa/doc/11.0/osa_um_ibis/Cookbook.html}}. For each scw, we create the sky model and reconstruct the sky image and the source count rates by deconvolving the shadowgrams projected onto the detector plane.

For the data analyzed with OSA 11.0, we extract spectra with 60 logarithmically spaced channels between 13\,keV and 1000\,keV. For the OSA 10.2 extraction, we create a response matrix with a binning that matches the one automatically generated by running the OSA 11.0 spectral extraction   as closely as possible. Response matrix channels differ at most by 0.25\,keV between the OSA 10.2 and OSA 11.0. We then use the OSA 10.2 \textit{spe\_pick} tool to create stacked spectra for each spectral state according to our state classification (see Sect. \ref{integral_class} below). We add 1.5\,\% of the systematic error to both OSA 10.2 and OSA 11.0 stacked spectra. The ISGRI spectra obtained with OSA 10.2 are analyzed in the 25--300\,keV energy range, while those obtained with OSA 11.0 are analyzed over the 30--300\,keV energy range. 

\subsection{\integral/SPI spectral extraction}
To reduce the data from the SPectrometer aboard \integral \citep[SPI;][]{Vedrenne2003} we use the SPI Data Analysis Interface\footnote{\url{https://sigma-2.cesr.fr/integral/spidai}.} to  extract averaged spectra. The sky model we create contains Cyg X-1 and Cyg X-3. We  typically set the source variability to ten scws for Cyg X-3 and five scws for Cyg X-1 \citep[e.g.,][]{Bouchet2003}. We then create the background model by setting the variability timescale of the normalization of the background pattern to ten scws. The background is generally stable, but solar flares, radiation belt entries, and other nonthermal incidents can lead to unreliable results. In order to avoid these effects, we remove scws for which the reconstructed counts compared to the detector counts give a poor $\chi^2$ ($\chi^2_\mathrm{red} > 1.5$). This selection reduces the total number of scws by $\sim 10$\,\%. The shadowgrams are deconvolved to obtain the source flux, and spectra are then extracted between 20\,keV and 400\,keV using 30 logarithmically spaced channels.

\section{State classification based on RXTE/PCA}
\label{classification}
\subsection{Proportional Counter Array data reduction and classification}

\begin{figure*}
\centering
\includegraphics[width=\textwidth]{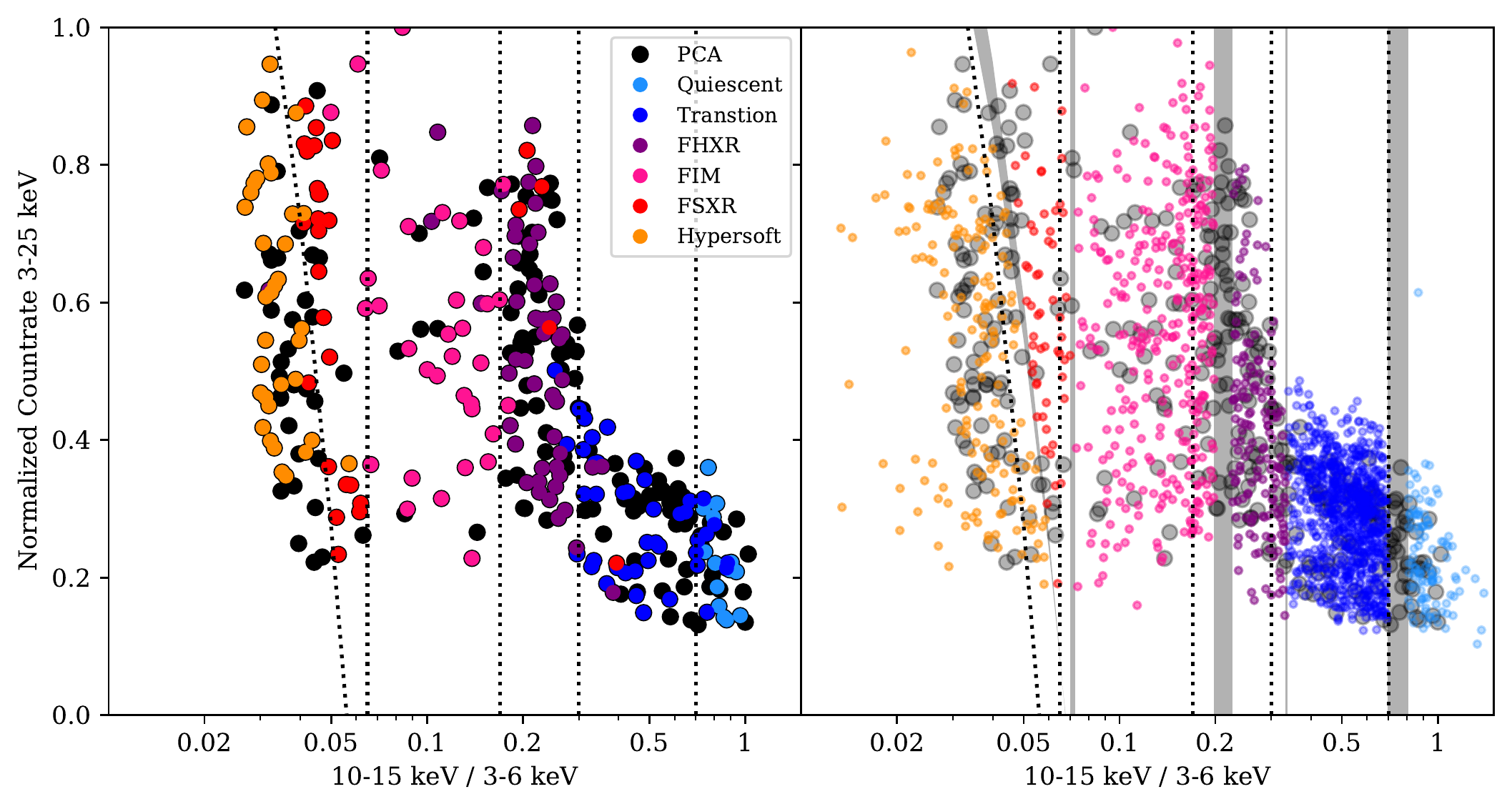}
\caption{HID of Cyg X-3. \textit{Left:} HID of Cyg X-3 using PCA data in black dots. Each colored dot corresponds to an observation which has  already been classified by K10. Vertical black-dashed lines correspond to our division into six states: Quiescent (light blue), Transition (dark blue), FHXR (purple), FIM (pink), FSXR (red), and hypersoft (orange). \textit{Right:} HID with PCA and JEM-X data superposed. Gray dots correspond to the same data as in the left plot and colored dots correspond to classified JEM-X scws using the same color code as in the left figure. Vertical gray bands correspond to the JEM-X state divisions using the method described in the text.}
\label{HID}
\end{figure*}

We consider all \textit{Rossi X-ray Timing Explorer} \citep[\textit{RXTE,}][]{Bradt1993} Proportional Counter Array \citep[PCA,][]{Glasser1994} \textit{Standard-2} observations of Cyg X-3, that is to say 262 observations from 1996 to 2011. This adds three years of new observations compared to the work done by K10. The data are reduced with the version 6.24 of the HEASOFT. We followed a procedure very similar to the one followed by \cite{Rodriguez2008, Rodriguez2008II} to filter out the data from bad time intervals and to obtain PCA light curves and spectra for GRS 1915+105, another peculiar and very variable microquasar. We consider the data from all layers of the  Proportional Counter Unit \#2, which is the best calibrated and the one that is always turned on. Background maps are estimated with \textit{pcabackest} using the bright model. Source and background spectra are obtained with \textit{saextrct} and response matrices by running the \textit{pcarsp} tool.

For all observations, we extract net source count rates using the \textit{show rates} command from \textsc{xspec} in three energy ranges:  3--6\,keV, 10--15\,keV, and 3--25\,keV. We use 3--6\,keV and 10--15\,keV in order to be as consistent as possible with the approach of K10, permitting us to probe the different spectral regions in a model-independent manner. The last energy band extends to 25\,keV in order to be consistent with the JEM-X spectral band that we use for our classification of the \integral data. Figure \ref{HID} (left) shows the HID obtained from the PCA data. Each point corresponds to one observation. The colored dots are the observations analyzed by K10, while the black ones are the new addition from the present work. PCA count rates are normalized according to the maximum count-rate value. The state of each observation is attributed according to Table A.1 of K10. We define the best state division based on this classification and divide the HID in six different zones.

\subsection{Extension of the \textit{RXTE/PCA} state classification to \integral}
\label{integral_class}

When evaluating the hardness ratio, the observed fluxes are convolved with the instrument response matrices, and thus are instrument dependent. Therefore, we cannot simply use the PCA state boundaries to classify the JEM-X data. The differences are illustrated in  Table \ref{annexe:simult} with 16 quasi-simultaneous JEM-X/PCA (i.e., within 0.1 MJD) observations. These examples show that there is no one-to-one correspondence between the JEM-X and PCA values of the HR. We therefore need to convert the PCA boundary values to those of JEM-X.  

To do so, and  to thus extend the classification of K10 to the JEM-X data, we proceed as follows. For each division (quiescent/transition, transition/FHXR, FHXR/FIM, FIM/FXSR, FSXR/hypersoft), we select the closest PCA observations from both sides of the state division line, and then we search the best functions that fit the data\footnote{We use the models \textsc{edge*edge*tbabs(cutoffpl + gaussian)} for the quiescent, transition and FHXR states, \textsc{edge*edge*tbabs(cutoffpl + gaussian + diskbb)} for the FIM state and \textsc{edge*edge*tbabs(powerlaw + gaussian + diskbb)} for FSXR and hypersoft states.} to the spectrum. Subsequently, we simulate JEM-X data using that model and the appropriate redistribution matrix which allows us to finally calculate the countrate of the same energy range as used in K10.

We use \textsc{xspec} version 12.9.1p for all data modeling \citep{Arnaud1996}. Thanks to the \textit{show rates}\footnote{This command gives the net count rate of the spectrum and model-predicted rate in the considered energy range.} command in \textsc{xspec}, we can find the corresponding hardness ratio in JEM-X. This allows us to draw new state divisions which can be used for JEM-X. Figure \ref{HID} (right) shows the JEM-X HID; the vertical gray bands correspond to the new JEM-X state divisions using the described method above. As for PCA, JEM-X count rates are normalized to the maximum value. For our accumulated JEM-X spectra, we only select scws that are outside the state division line (i.e., colored dot in the plot) in order to be sure that our accumulated spectra are not polluted by scws with different spectral classification. After this selection, a total of 1501 classified scws remain.

As the SPI determination background is based on the dithering pattern \citep{Vedrenne2003}, we have to select continuous sets of scws ($\sim 15$) in order to obtain good precision on the flux evaluation. The JEM-X field of view is smaller than that of SPI or IBIS. It appears that for some observations, the source is outside the JEM-X field of view while being observed by SPI and/or IBIS. When considering a set of $> 15$ scws, some of them remain unclassified, as we build our classification with JEM-X observations. To try and overcome this problem in order to exploit SPI data, we specially create new lists of scws for the SPI analysis. To obtain lists of 15 consecutive classified scws, we make the following approximation: we consider that a given SPI scw has the same state as the previously (or next) JEM-X classified one if (1) the classification of the source is known for scws distant from < 10 scws, (i.e., the variability threshold of the source) and (2) no obvious state change is seen in the global light curve. 

As the list of classified scws in the FSXR state does not verify these conditions, we do not extract a SPI spectrum for this state. Table \ref{nb_scws} summarizes the number of scws for the different extractions.

\begin{table}[h]
\tiny{
\centering
\caption{Number of scws for each spectral state.}
\label{nb_scws}

\begin{tabular}{l c c c c c} \hline
State & JEM-X 1 & JEM-X 2 & ISGRI$_\mathrm{OSA 10}$ & ISGRI$_\mathrm{OSA 11}$ & SPI \\ \hline 
Quiescent & 39 & 63 & 91 & 11& 47 \\
Transition & 648 & 59 & 653 & 54 & 559\\
FHXR & 134 & 9 & 143 & 0 & 96\\
FIM & 314 & 0 & 313 & 1 & 475\\
FSXR & 54 & 0 & 54 & 0 & 0\\
Hypersoft & 176 & 5 & 168 & 13 & 223\\ \hline
\end{tabular}
}
\end{table}

\subsection{Images}

\begin{figure*}[h]
\centering
\includegraphics[width=\textwidth]{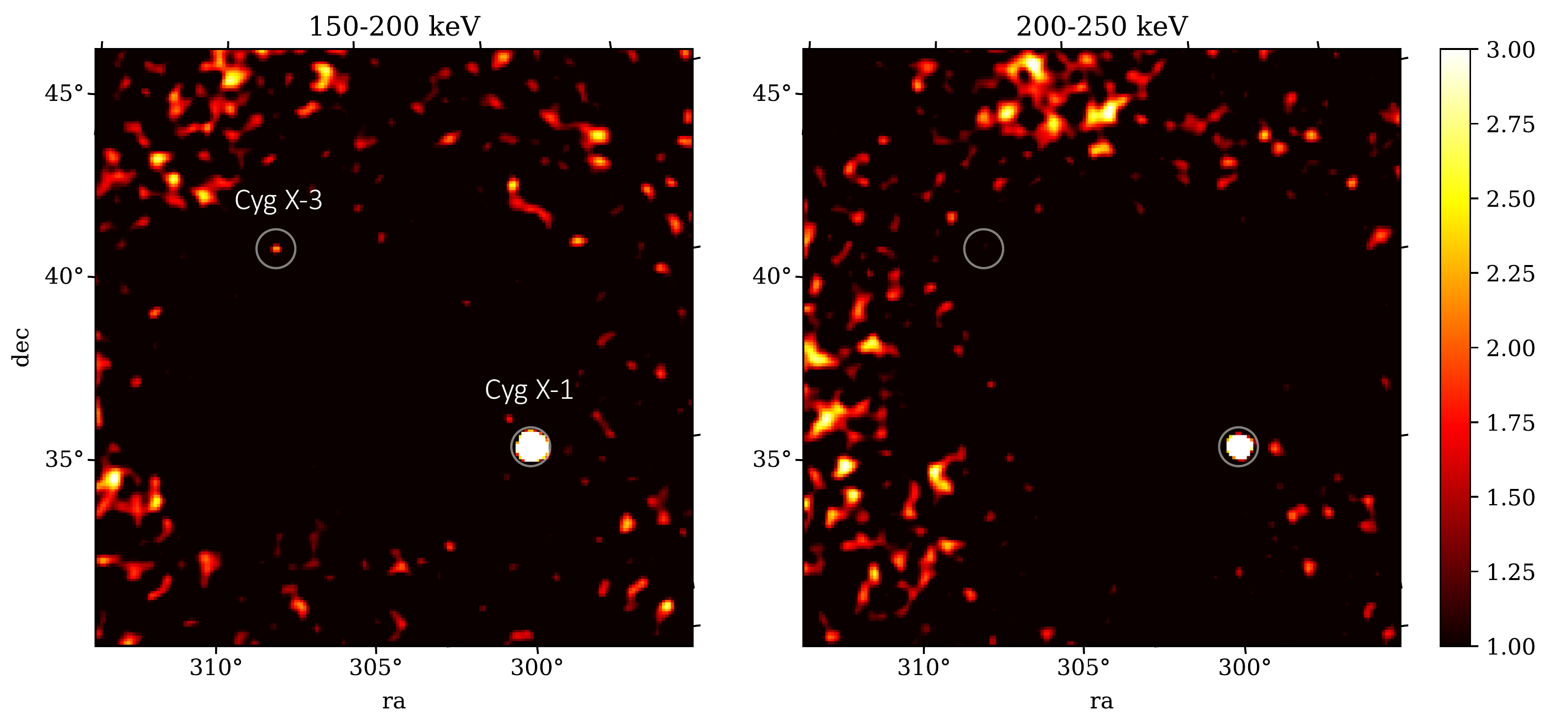}
\caption{IBIS stacked images of the Cygnus region in two energy bands, 150--200\,keV (left) and 200--250\,keV (right), when considering the whole data set.}
\label{fig:images}
\end{figure*}

We extract stacked images with IBIS  in three energy bands: 150--200\,keV, 180--200 keV, and 200--250 keV. Figure \ref{fig:images} shows images of the Cygnus region in the 150--200\,keV (left) and 200--250\,keV (right) energy bands when considering the whole data set. Cyg X-3 is detected in the 150--200\,keV band with a significance of 7.65 and is not detected in the 180--220\,keV or 200--250\,keV bands.

When considering both quiescent and transition states, Cyg X-3 is detected in the 150--200\,keV range with a significance of 7.02. We find a hint of detection at 2.54$\sigma$ in the 180--220\,keV band. Above 200\,keV it is not detected. Finally, when considering flaring (FHXR, FIM and FSXR) and hypersoft state, the source is detected in the 150--200\,keV band with a significance of 3.89.

\section{Spectral fitting: phenomenological approach}
\label{phenom}

\begin{figure*}[t]
\centering
\includegraphics[width=\textwidth]{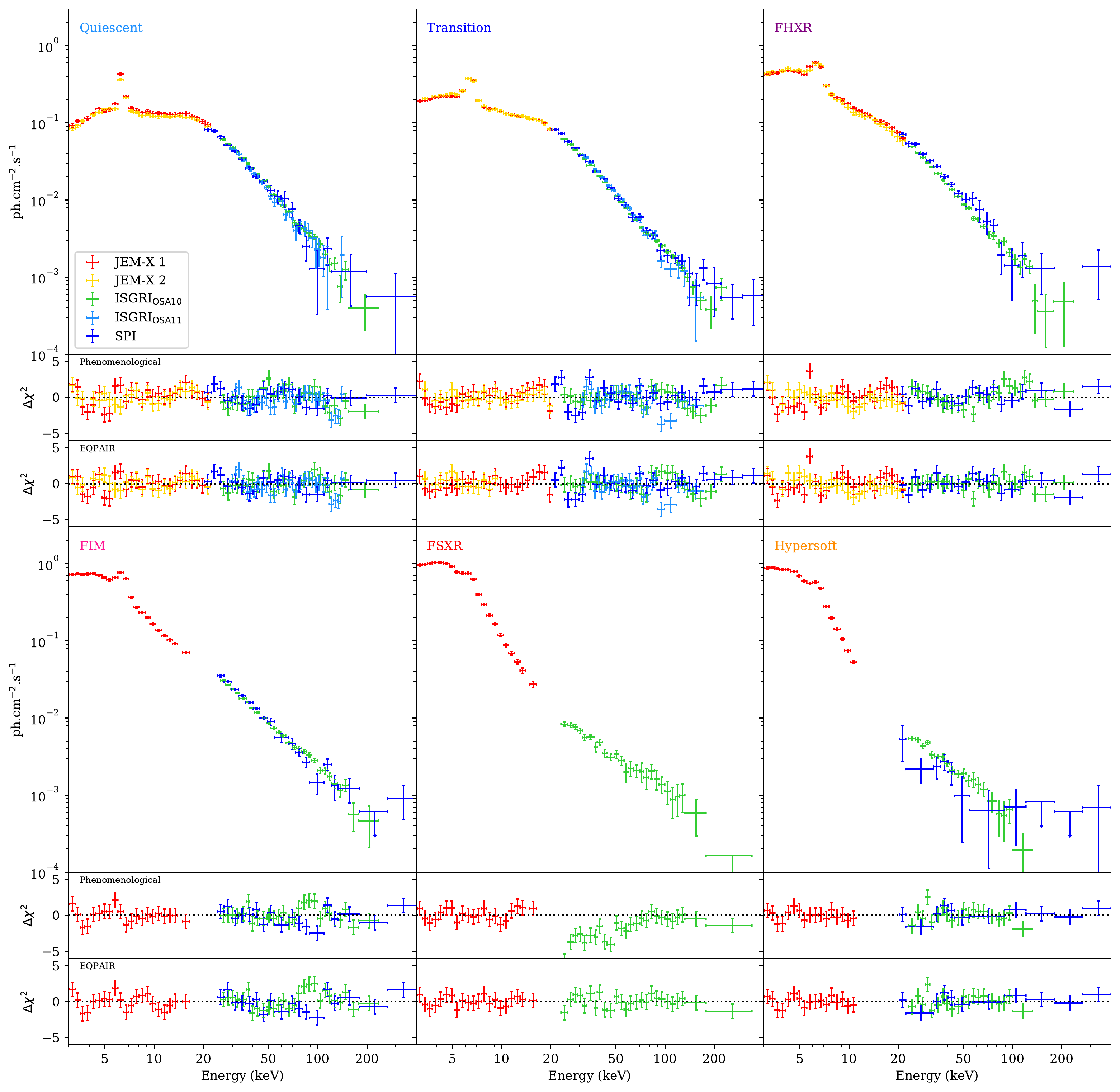}
\caption{Accumulated state-resolved spectra for each state according to our classification. Residuals of our phenomenological and physical (\eqpair) fitting are shown.}
\label{all_spec}
\end{figure*}

Because of the poor statitistics of the JEM-X 2 and ISGRI/OSA 11.0 spectra in the FIM and hypersoft states, we do not use them for our spectral fitting. Therefore, we use the five spectra (JEM-X 1, JEM-X 2, ISGRI/OSA 10.2, ISGRI/OSA 11.0 and SPI) for the quiescent and transition states, four (JEM-X 1, JEM-X 2, ISGRI/OSA 10.2 and SPI) for the FHXR state, three (JEM-X 1, ISGRI/OSA 10.2 and SPI) for the FIM and hypersoft states, and two (JEM-X 1 and ISGRI/OSA 10.2) for the FSXR. Figure \ref{all_spec} shows the obtained number of individual spectra that were stacked and fitted simultaneously.

We first use a phenomenological approach to investigate the spectral behavior of the source at high-energy and test the possible presence of a high-energy tail above $\sim 100$\,keV.

\subsection{Method}
\label{phenom_approach}

We first model our spectra with an absorbed \citep[\textsc{tbabs},][]{Wilms2000} power law and an iron line in \textsc{xspec}. The iron line centroid is fixed to 6.4\,keV and its width is limited below 0.4\,keV. We use \textsc{angr} solar abundances \citep{Anders1982}. A simple power law does not provide acceptable fits to the quiescent, transition, or FHXR spectra. Large residuals around 20\,keV indicate the presence of a break or a cut-off, in agreement with previous findings (K10, H09). We then use a powerlaw with cutoff \textsc{cutoffpl} instead. To obtain statistically good fits, a reflection component \cite[\textsc{reflect},][]{Zdziarski1995} is also added in the quiescent, transition, FHXR, and FIM states. The reflection factor is limited to below 2. We also add a multicolor black-body component (\textsc{diskbb}) for the FIM, FSXR, and hypersoft states. 

Without the addition of a power law in the quiescent and transition states, we obtain significant residuals at high energy. The reduced $\chi^2$ values are 5.0 (119 dof) and 8.3 (121 dof), respectively. We test the significance of this additional component by performing an F-test. We find F-test probabilities (i.e., that the statistical improvement due to the addition of the new power-law component is due to chance) of $7.9 \times 10^{-33}$ and $2.4 \times 10^{-47}$ for the quiescent and the transition states respectively.
 In summary, we use: \textsc{constant*tbabs*reflect(cutoffpl + powerlaw + gaussian)} for quiescent and transition states, \textsc{constant*tbabs*reflect(cutoffpl + gaussian)} for the FHXR state, and \textsc{constant*tbabs(powerlaw + diskbb + gaussian)} for FIM, FSXR, and hypersoft states. The constant component allows us to take into account calibration issues between different instruments and the different sample of scws to build instrument spectra. We fix $i = 30^\circ$ as inclination for the reflection \citep{Vilhu2009, Zdziarski2012, Zdziarski2013}.
The 3--300\,keV spectral parameters obtained for each spectral state are reported in Table \ref{tab:param_3-300}.

\begin{table*}[h]
                        \center
                        \caption{Parameters for our phenomenological fitting for each state. Fixed parameters are indicated with an \textquotedblleft f\textquotedblright.}
                        
                        \label{tab:param_3-300}
                        \small{
                        \begin{tabular}{l c c c c c c c c}
                        \hline
                        Parameters & Quiescent &  Transition & FHXR & FIM & FSXR & Hypersoft\\
                        \hline
                        $C_\mathrm{JEMX-1}$ & 1\,f & 1\,f & 1\,f& 1\,f &  1\,f & 1\,f  \\ [3pt]
                        
                        $C_\mathrm{JEMX-2}$ &  $0.94^{+0.02}_{-0.02}$& $1.01^{+0.02}_{-0.02}$& $0.98^{+0.02}_{-0.02}$& -- & -- & --\\ [3pt]
                        
                        $C_\mathrm{OSA 10.2}$ & $0.85^{+0.02}_{-0.02}$ & $0.92^{+0.02}_{-0.02}$& $0.90^{+0.03}_{-0.01}$ & $1.18^{+0.02}_{-0.04}$ & $0.80^{+0.01}_{-0.00}$& $1.19^{+0.01}_{-0.14}$\\ [3pt]
                        
                        $C_\mathrm{OSA 11.0}$ & $0.80^{+0.02}_{-0.00}$ &$1.02^{+0.03}_{-0.03}$ & -- & -- & -- & --\\ [3pt]
                        
                        $C_\mathrm{SPI}$ & $0.90^{+0.02}_{-0.02}$ & $0.96^{+0.02}_{-0.02}$ & $0.99^{+0.01}_{-0.03}$& $1.23^{+0.02}_{-0.02}$& -- & $1.03^{+0.01}_{-0.14}$\\ [3pt]
                        
                        $N_\mathrm{H} \times 10^{22}$ [cm$^{-2}$] & $8.7^{+1.3}_{-1.3}$ & $7.29^{+0.82}_{-0.32}$ & $10.53^{+0.59}_{-0.23}$ & $8.16^{+0.25}_{-0.25}$ & $6.4^{+0.9}_{-1.0}$ & $5.07^{+0.97}_{-0.84}$ \\ [3pt]
                        
                        $\Gamma_\mathrm{cut}$ & $0.97^{+0.25}_{-0.31}$ & $1.457^{+0.009}_{-0.38}$ & $3.059^{+0.058}_{-0.007}$& -- & -- & --  \\ [3pt]
                        
                        $kT_\mathrm{disk}$ [keV] & -- & -- & -- & $1.25^{+0.01}_{-0.01}$& $1.36^{+0.04}_{-0.04}$& $1.301^{+0.009}_{-0.037}$\\ [3pt]

                                $E_\mathrm{cut}$ [keV] & $14.1^{+1.9}_{-1.1}$ & $14.2^{+1.6}_{-0.2}$ & $434^{+157}_{-93}$ & -- & -- & -- \\ [3pt]

                        $\sigma_\mathrm{Fe}$ [keV] & $0.26^{+0.14}_{-0.01}$ & $0.35^{+0.19}_{-0.05}$ & $0.40^{+0.00}_{-0.23}$ &  $0.40^{+0.00}_{-0.09}$  & 0.4\,f & 0.4\,f\\ [3pt]
                        
                        $\Omega/2\pi$ & $1.25^{+0.52}_{-0.41}$ & $1.99^{+0.01}_{-0.44}$ & $1.99^{+0.01}_{-0.16}$ & $0.24^{+0.22}_{-0.12}$& -- & -- \\ [3pt]
                        
                        $\Gamma_\mathrm{po}$ & $2.21^{+0.11}_{-0.14}$ & $2.571^{+0.004}_{-0.10}$ & -- & $2.917^{+0.006}_{-0.006}$& $2.50^{+0.10}_{-0.11}$ & $2.68^{+0.11}_{-0.12}$ \\ [3pt]
                        
\hline
Total flux 50--300\,keV & & & & & &  \\ [3pt]
[$10^{-10}$ erg/s/cm$^{2}$] & 7.4 & 6.6 & 4.7 & 6.4 & 7.9 & 5.7\\ [3pt]

\hline
Total flux 3--300\,keV & & & & & & \\ [3pt]
[$10^{-9}$ erg/s/cm$^{2}$] & 6.7 & 6.7 & 7.5 & 8.4 & 2.9 & 1.4 \\ [3pt]

                        $\chi^2/\mathrm{dof}$ & 189.57/120 & 165.27/119 & 128.45/91 & 84.04/59 & 30.32/41 & 33.50/42\\ [3pt]
                        \hline
                        \end{tabular}
                        
                        }
                        \end{table*}

\subsection{Results}
The best parameters obtained in the 3--400\,keV band are reported in Table \ref{tab:param_3-300}, and Fig. \ref{all_spec} shows the spectra and best-fit models for each state.
The value of the column density varies between 5 and 10 $\times 10^{22}$\,cm$^{-2}$ with a mean value of $\sim 7.6 \times 10^{22}$\,cm$^{-2}$. For the quiescent and transition states, the cutoff energy is around 15\,keV, matching with previous results (K10, H09). For the FHXR, the cutoff value is quite high ($E_\mathrm{cut} = 434$\,keV). We therefore, try to use a simple power law to represent this spectrum, but this model does not converge to an acceptable fit (with a reduced $\chi^2$ of 738.51/93 dof), showing the need for the cutoff albeit with poorly constrained parameters. The increase in cutoff energy is correlated with the increase in the photon index of the cutoff power law which evolves from $\Gamma_\mathrm{cut} \sim 1.5$ in the transition state to $\Gamma_\mathrm{cut} \sim 3$ in the FHXR state. The disk temperature $kT_\mathrm{disk}$ found in the FIM, FSXR, and hypersoft states is consistent with the results of K10 and H09.

Concerning the simple power law (additional or not), we observe a photon index of $\Gamma_\mathrm{po} \sim 2.5$ with a broadly similar value in three out of the six states (transition, FSXR, hypersoft). The FHXR does not show any power-law component. In the quiescent state, $\Gamma_\mathrm{po}$ is marginally compatible with these, reaching $2.21^{+0.11}_{-0.14}$, its lowest value of all states. In the FIM, $\Gamma_\mathrm{po}$ reaches its highest value ($2.917 \pm 0.006$) and clearly shows a different value compared to the other states (the average value of the photon index in the other state is \textlangle\,$\Gamma$\,\textrangle $= 2.49^{+0.18}_{-0.23}$).

In order to verify that our results, and in particular the need for an extra component, do not depend too strongly on the curved low-energy ($<50$\,keV) spectra, we investigate the 30--200 keV range alone. While in five out of six of the states, a simple power law fits the spectra well, a clear broken power law with an energy break at $73 \pm 8$\,keV is needed for the transition state (the one with the highest statistics). We find $\Gamma_1 = 3.52 \pm 0.03$ and $\Gamma_2 = 3.20 \pm 0.10$. This confirms the existence of an additional component to the simple power law above typically a few tens of keV. 

\subsection{Influence of the orbital modulation?}

\begin{figure}[h]
\centering
\includegraphics[width=\columnwidth]{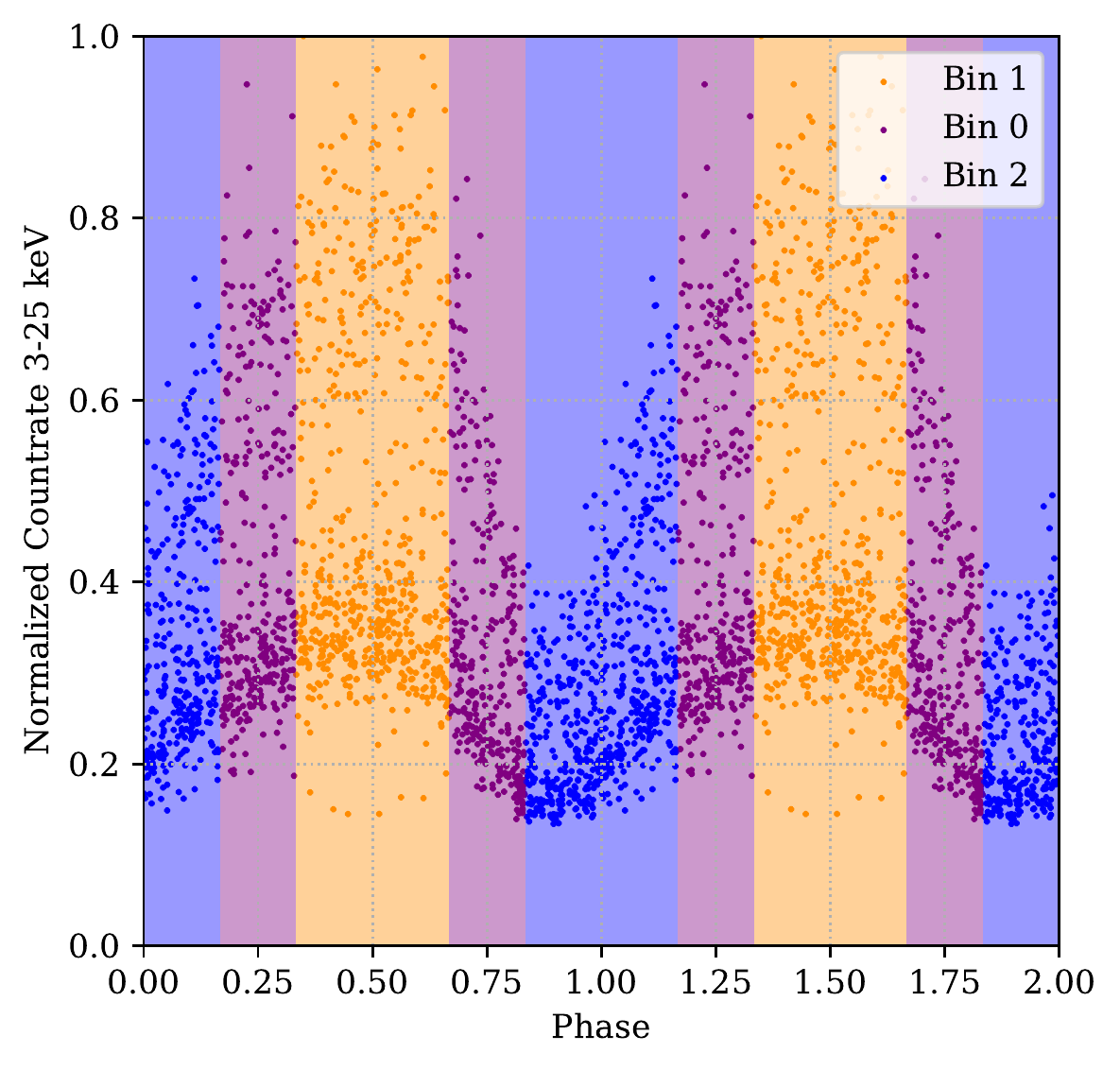}
\caption{Orbital light curve of Cyg X-3; each point corresponds to a scw. The three different bins are represented in blue (bin 1: inferior conjunction $1/3 < \phi < 2/3$), orange (bin 2: superior conjunction $0 < \phi < 1/6$ and $5/6 < \phi < 1$), and purple (bin 0: between the two conjunctions).}
\label{orbital_phase}
\end{figure}

Cyg X-3 shows strong orbital modulation in its X-ray flux \citep[e.g.,][]{Zdziarski2012}. We therefore investigate whether the presence of the tails is correlated with the orbital modulation. In order to do this, we use the ephemeris of \cite{Singh2002} and create a phase-binned light curve. We define three different phase bins: the first corresponds to the inferior conjunction, that is, $1/3 < \phi < 2/3$ (the compact object being in front of the star), the second corresponds to the superior conjunction, that is, $0 < \phi < 1/6$ and $5/6 < \phi < 1$ (the compact object being behind the star), and the third bin corresponds to the transition between the two others. Figure \ref{orbital_phase} shows the folded light curve; each point represents one scw, and scw is classified according to a phase bin: `1' for inferior conjunction, `2' for superior conjunction, and `0' between the two conjunctions.

\begin{figure*}[h]
\centering
\includegraphics[width=\textwidth]{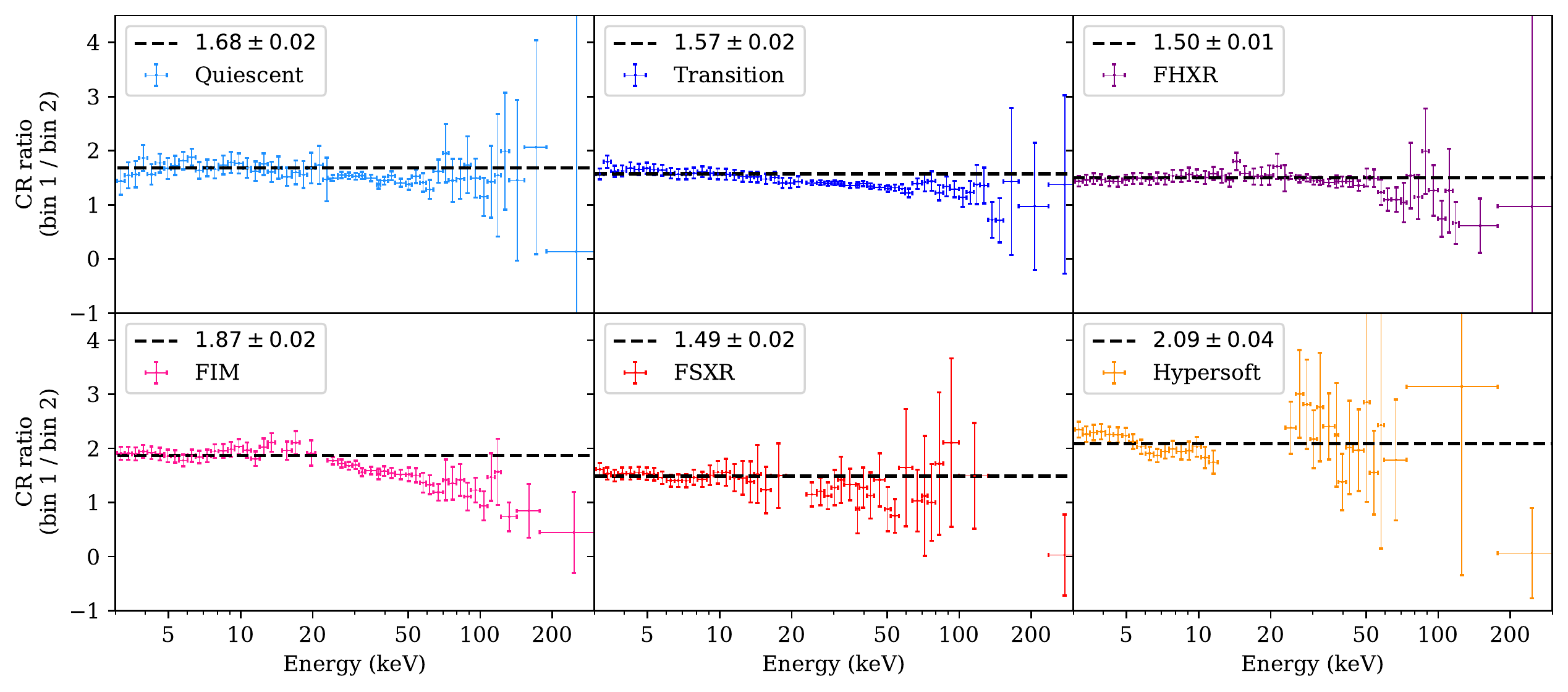}
\caption{Count rate ratio between the spectra of bin 1  and bin 2 for each state according to the same color code as in Fig. \ref{HID}.}
\label{ratio}
\end{figure*}

To check on the potential influence of the orbit, we check the two extreme positions and create stacked orbital-phase- and state-dependent spectra. Figure \ref{ratio} shows the count-rate ratio between spectra extracted from bin 1 (inferior conjunction) and spectra extracted from bin 2 (superior conjunction). As the ratio remains at a constant value of $\simeq 2$ from 3 to 20 keV, we observe a slight decrease starting from 20 keV to 300 keV for the transition, FHXR, and FIM states which are not statistically significant in the other states. Dashed lines represent the best fits with a constant in the energy range 3--20\,keV; results are indicated in the legend of Fig. \ref{ratio}.

We investigate the possible change in the slope of the spectrum as a function of the phase bin. In order to do this, we use the same phenomenological model as described in Sect. \ref{phenom_approach}. The values of the photon index are reported in Table \ref{orbit_param} for each state and for each bin. 

The values are compatible in their 90\,\% confidence range, excepted in the FIM state where we observed an increase of 7.5\,\% of the photon index value when Cyg X-3 is in inferior conjunction, that is, the spectrum is softer when the source is in front of the star. This behavior is discussed in Sect. \ref{sec:orbital_mod}.

\begin{table}[h]

\caption{Measured values of the photon index for each state and for each phase bin based on the phenomenological model from Sect. \ref{phenom_approach}.}
\label{orbit_param}
\begin{center}
\begin{tabular}{l c c c}
\hline
State &  & Bin 1 & Bin 2 \\
\hline
Quiescent & $\Gamma_\mathrm{po}$ & $2.45^{+0.17}_{-0.18}$ & $2.16^{+0.17}_{-0.26}$ \\ [3pt]
Transition & $\Gamma_\mathrm{po}$ & $2.55^{+0.05}_{-0.05}$ & $2.45^{+0.08}_{-0.20}$ \\ [3pt]
FHXR & $\Gamma_\mathrm{cut}$ & $2.90^{+0.06}_{-0.20}$ & $3.00^{+0.02}_{-0.04}$ \\ [3pt]
FIM & $\Gamma_\mathrm{po}$ & $2.94^{+0.04}_{-0.04}$ & $2.72^{+0.04}_{-0.04}$ \\ [3pt]
FSXR & $\Gamma_\mathrm{po}$ & $2.52^{+0.16}_{-0.16}$ & $2.47^{+0.24}_{-0.17}$ \\ [3pt]
Hypersoft & $\Gamma_\mathrm{po}$ & $2.87^{+0.18}_{-0.16}$ & $2.59^{+0.33}_{-0.28}$ \\ [3pt]
\hline
\end{tabular}
\end{center}
\end{table}

\section{Physical approach: hybrid thermal/nonthermal model: \textsc{eqpair}}
\label{sec:eqpair}

In order to better constrain properties of the nonthermal component observed in our phenomenological approach, we  now apply a more physical model of hybrid thermal/nonthermal coronae to our spectra.

\subsection{The \textsc{eqpair} model}
The complete model is described in \cite{Coppi1999}; here we give a brief summary. In this model, the total luminosity of the source $L_\mathrm{rad}$ is re-expressed as a dimensionless parameter, the \textquotedblleft compactness\textquotedblright\,$l_\mathrm{rad}$: 

\begin{equation}
 l_\mathrm{rad} = \frac{L_{\mathrm{rad}}}{R} \frac{\sigma_\mathrm{T}}{m_ec^2}
 ,\end{equation}
where $R$  is the characteristic radius of the corona (assuming it is spherical), $\sigma_\mathrm{T}$ is the Thomson cross-section, $m_e$ the electron mass, and $c$ the speed of light. The luminosity from soft photons from the disk is parametrized by another compactness parameter $l_\mathrm{s}$, and the spectrum shape of these soft photons is assumed to be a black body with a temperature $kT_\mathrm{bb}$. The amount of heating is expressed by the ratio of the compactness of the Comptonized medium and the compactness of the seed photons $l_\mathrm{h}/l_\mathrm{s}$. In this model, electrons from a cool background plasma with an optical depth $\tau_\mathrm{p}$ are accelerated to form the observed nonthermal tail. Thus, we assume the Lorentz factor of the accelerated nonthermal plasma to be distributed according to a  power law within the range $\gamma = 1.3$--1000.

The luminosity of these nonthermal electrons $L_\mathrm{nth}$ is once again described by the dimensionless compactness $l_\mathrm{nth}$. Nonthermal processes from where particles are allowed to cool are Compton scattering, synchrotron radiation, and Bremsstrahlung emission. To balance these nonthermal processes, the compactness $l_\mathrm{th}$ represents the dimensionless luminosity from thermal interaction between particles, i.e., Coulomb interaction. The reflection model implemented is \textsc{ireflect} \citep{Zdziarski1995}.

For our fitting with this hybrid model, we add absorption and an iron line, and therefore the model is computed in \textsc{xspec} as:
\begin{equation*}
\mathrm{\textsc{constant*tbabs(eqpair + gaussian).}}
\end{equation*}
Exceptionally, we need to add an ionized iron edge to correctly describe the FIM state.
The parameters allowed to vary freely in \textsc{xspec} are the ratio $l_\mathrm{h}/l_\mathrm{s}$, which is related to the slope of the Comptonizing spectrum, $l_\mathrm{nth}/l_\mathrm{h}$, the temperature of the black-body $kT_\mathrm{bb}$, the optical depth $\tau_p$, the index of the injected electrons distribution $\Gamma_\mathrm{inj}$, the fraction of the scattering region intercepted by reflecting material $\Omega/2\pi,$ and the width of the iron line (restricted to  a maximum value of 0.4\,keV). 

The luminosity from the seed photons is not well constrained, but the $\chi^2$ minimum oscillates between $l_\mathrm{s} = 40$ and 140, and therefore we fix it at the value of $l_\mathrm{s} = 100$, which correspond to a small radius of the corona for a high luminosity \citep{Zdziarski2005}.

\subsection{Results}

Table \ref{physical} summarizes the parameters obtained with the hybrid model and Fig. \ref{all_spec} shows the residuals for each state. We observe that the quiescent and transition states are characterized by a high value of $l_\mathrm{nth}/l_\mathrm{h}$ ($> 60$\,\%) and $l_\mathrm{h}/l_\mathrm{s}$ ($\gg 1$) expressing a spectrum dominated by Comptonization processes. The energy cutoff around 15--20\,keV and the shape of the nonthermal component with an electron injection index of $\sim 3.5$ are well reproduced by the model. The reflection parameter is rather small compared to our phenomenological approach.

In the FHXR state, the value of $l_\mathrm{h}/l_\mathrm{s} = 0.75$ decreases significatively. We observe a clear rise in the photon temperature $kT_\mathrm{s}  = 390$\,eV. We find a value for reflection of $\Omega/2\pi = 0.94^{+0.13}_{0.10}$ compatible with the one found in quiescent and transition states, and the electron injection index $\Gamma_\mathrm{inj} = 3.57^{+0.41}_{-0.15}$ remains close to the values found previously. 

Concerning the FIM state, the spectrum is characterized by an important photon disk compacity ($l_\mathrm{h}/l_\mathrm{s} = 0.51 \pm 0.03$ i.e., $\sim$ 50\,\% of the total luminosity is supplied by the disk emission). The fraction $l_\mathrm{nth}/l_\mathrm{h} = 0.50^{+0.11}_{-0.10}$ and the electron injection index $\Gamma_\mathrm{inj} = 2.78 \pm 0.14$  are also smaller than in the FHXR state. This state is very close to the \textit{very high state} of H09.

Finally, the FSXR and  hypersoft states are described by an important contribution of the disk photons ($l_\mathrm{h}/l_\mathrm{s} \sim 0.1$ i.e., $\sim$90\,\% of the total luminosity is supplied by the disk emission). They also show an important nonthermal emission ($l_\mathrm{nth}/l_\mathrm{h} > 50$\,\% i.e., the total heating is dominated by nonthermal processes) with an electron injection index of $\Gamma_\mathrm{inj} \sim 3$. The values of $\tau_\mathrm{p}$ are not well constrained in those states, and we remark that they, in particular, are not consistent with the results of the \textit{nonthermal} and \textit{ultrasoft state} found by H09. Moreover, the electron injection index in their ultrasoft state is smaller, resulting in a much harder spectrum at higher energy.

\begin{table*}[t]

\small{
\caption{Parameters for the \textsc{eqpair} fitting. Fixed parameters are indicated with an \textquotedblleft f\textquotedblright. The electron temperature $k\mathrm{T_e}$ is calculated from the energy equilibrium, i.e., not a free fit or a fixed parameter.}
\label{physical}
\begin{center}
\begin{tabular}{l c c c c c c}
\hline 
State & Quiescent &  Transition & FHXR &  FIM & FSXR &  Hypersoft\\ [3 pt]
\hline

$C_\mathrm{JEMX-1}$ & 1\,f &  1\,f &  1\,f  & 1\,f  &  1\,f  &  1\,f  \\ [3 pt]
                                                $C_\mathrm{JEMX-2}$ & $0.93^{+0.02}_{-0.02}$ & $0.95^{+0.03}_{-0.03}$ & $0.99^{+0.02}_{-0.02}$ & -- &  -- & -- \\ [3 pt]

                                                $C_\mathrm{ISGRI_\mathrm{O10}}$ & $0.85^{+0.02}_{-0.03}$ & $1.01^{+0.02}_{-0.02}$ & $0.94^{+0.04}_{-0.04}$ & $1.07^{+0.02}_{-0.06}$ & $0.80^{+0.03}_{-0.00}$ & $0.83^{+0.03}_{-0.00}$ \\ [3 pt]

                                                $C_\mathrm{ISGRI_\mathrm{O11}}$& $0.80^{+0.03}_{-0.00}$ & $1.02^{+0.03}_{-0.03}$ & -- & -- & -- & --\\ [3 pt]
                                                
                                                $C_\mathrm{SPI}$& $0.92^{+0.02}_{-0.03}$ & $1.00^{+0.02}_{-0.01}$ & $1.04^{+0.02}_{-0.01}$ & $1.07^{+0.02}_{-0.02}$ & -- & $1.20^{+0.00}_{-0.13}$\\ [3 pt]

                                                $N_\mathrm{H}$ [$\times 10^{22}$\,cm$^{-2}$] & $6.75^{+0.78}_{-0.30}$  & $5.48^{+0.67}_{-0.33}$ & $8.8^{+0.6}_{-1.1}$ & $10.2^{+1.3}_{-1.5}$ & $6.2^{+1.1}_{-2.0}$ & $3.2^{+1.5}_{-1.5}$ \\ [3 pt]

                                                $l_\mathrm{h}/l_\mathrm{s}$ & $39.2^{+5.0}_{-6.6}$ & $5.8^{+2.6}_{-2.7}$ & $0.75^{+0.09}_{-0.15}$& $0.51^{+0.03}_{-0.03}$ & $0.15^{+0.16}_{-0.05}$ & $0.11^{+0.02}_{-0.02}$ \\ [3 pt]

                                                $kT_\mathrm{s}$\,[eV] & $1.91^{+0.91}_{-0.67}$ & $4^{+16}_{-1}$ & $390^{+19}_{-49}$& $651^{+125}_{-258}$ & $987^{+48}_{-187} $ & $893^{+191}_{-115}$ \\ [3 pt]

                                                $l_\mathrm{nth}/l_\mathrm{h}$ & $> 0.92$ & $0.63^{+0.03}_{-0.03}$& $0.77^{+0.07}_{-0.16}$ & $0.50^{+0.11}_{-0.10}$ & $0.72^{+0.18}_{-0.03}$ & $0.51^{+0.44}_{-0.03}$\\ [3 pt]

                                                $\tau_\mathrm{p}$ & $< 5.1$ & $6.44^{+0.11}_{-0.21}$ & $5.6^{+1.3}_{-1.3}$ & $3.37^{+0.39}_{-0.39}$ & $< 0.86$ & $0.5^{+3.2}_{-0.4}$\\ [3 pt]

                                                $\Gamma_\mathrm{inj}$ & $3.60^{+0.14}_{-0.05}$ & $3.31^{+0.14}_{-0.09}$& $3.57^{+0.10}_{-0.06}$ & $2.78^{+0.14}_{-0.14}$ & $2.82^{+0.47}_{-0.06}$ & $3.15^{+0.47}_{-0.80}$\\ [3 pt]

$\Omega/2\pi$ & $0.80^{+0.26}_{-0.17}$ & $0.67^{+0.25}_{-0.11}$ & $0.94^{+0.13}_{-0.10}$ & $0.78^{+0.30}_{-0.45}$ & $ < 0.12$ & $<0.74$ \\ [3 pt]

$E_\mathrm{Fe}$ [keV] & 6.4\,f & 6.4\,f & 6.4\,f & 6.4\,f & 6.4\,f & 6.4\,f \\ [3 pt]

$\sigma_\mathrm{Fe}$[keV] & $0.39^{+0.01}_{-0.14}$ & $0.40^{+0.00}_{-0.13}$ & $0.40^{+0.00}_{-0.23}$ &  $0.39^{+0.01}_{-0.03}$& $0.40^{+0.00}_{-0.10}$ & 0.4\,f \\ [3 pt]

$E_\mathrm{E}$ [keV] &-- & -- & -- & $9.60^{+0.56}_{-0.47}$ & --  & -- \\ [3 pt]
$\tau_\mathrm{E}$ &-- & -- & -- & $0.24^{+0.10}_{-0.09}$ & --  & -- \\ [3 pt]

$k\mathrm{T}_\mathrm{e}$ [keV] & 4.04 & 4.25 & 3.42 & 5.76 & 5.75 & 3.36 \\ [3 pt]

$\chi^2/\mathrm{dof}$ & $123.89/116$ & $141.55/118$& $99.44/88$& $65.07/55$ & $25.82/37$ & $29.82/38$ \\ [3 pt]

\hline
\end{tabular}
\end{center}
}

\end{table*}

\section{Discussion}
\label{discussion}

We use the whole \integral database of Cyg X-3 in order to extract stacked spectra for each state previously defined by K10. Although this static approach is not adapted to study the source variability, it permits us to obtain results at high energies ($> 100$\,keV) that  are more statistically robust than in all previous studies, allowing us to probe the properties of the nonthermal hard-X-ray emission with the highest sensitivity. 

\subsection{Origin of the high-energy tail}
It is interesting to note that a nonthermal power-law-like component is present in all the states of Cyg X-3. The differences in the photon indices of this detected tail can be explained by one of two main scenarios: (1) the mechanism that gives rise to the tail is the same in all the states and endures some changes during state transitions, or (2) the mechanism is different depending on the state.

(1) All of our spectra are statistically well modeled by the thermal/nonthermal corona model \eqpair. With this model, the power-law component observed comes from a nonthermal distribution of electrons. The differences observed in the electron injection indices, and especially between those of the quiescent/transition/FHXR and FIM/FSXR states seem to point to a modification of the mechanism responsible for the electron acceleration through state transition. 

We know that in the quiescent and transition states, the radio flux observed is 60--300\,mJy (K10) and the radio spectrum is flat, implying the presence of compact jets in those states. On the other hand, we also know that powerful ejections with a radio flux of about 10\,Jy \citep{Szostek2008, Corbel2012} take place in the FIM state immediately after a period in the hypersoft state where the radio flux is quenched. This could indicate that the mechanism responsible for the electron acceleration in the corona is linked to the  behavior of the jet (compact jet vs. discrete ejections). This link has also been observed by \cite{Szostek2008} and \cite{Corbel2012} where these authors see a correlation between the radio flux and the hard-X-ray flux (30--80\,keV) during a major outburst. In this scenario, the nonthermal Comptonization component varies less than the thermal one, which is in turn responsible for the large variations of the high-energy flux. This would be compatible with the interpretation of the thermal corona being the base of the compact jet \citep{Markoff2005}, because the latter is also seen to vary greatly (disappear) as the source transits from the hardest states to the softest.

(2) Even if a single hybrid Comptonization represents all the data well, it is also
possible that the high-energy tail has a different origin depending on the states,
especially in states where the high energies are dominated by thermal Comptonization
while a compact jet is seen in radio (K10). The direct influence of the synchrotron emission from a hard state jet has been proposed in the case of Cygnus X-1 \citep{Laurent2011, Jourdain2014, Rodriguez2015}, another high-mass microquasar, while hybrid corona could be at the origin of the high-energy emission in softer states (Cangemi et al. submitted). To carry out a basic test of this possibility in Cyg X-3 we gather infrared data from \cite{Fender1996} and a radio spectrum from \cite{Zdziarski2016}. The infrared data were collected using the \textit{United Kingdom Infra-Red Telescope} (\textit{UKIRT}) on August 7, 1984, when the source was in its quiescent state. Fluxes are dereddened using the $\lambda^{-1.7}$ extinction law of \cite{Mathis1990}, and, following \cite{Fender1996}, we use $A_\mathrm{j} = 6.0$ as the extinction value. A radio spectrum from \cite{Zdziarski2016} was obtained by averaging the \textquotedblleft hard state\textquotedblright\,data from three measurements at 2.25, 8.3 (Green Bank Interferometer monitoring from November 1996 and October 2000), and 15 GHz (Ryle telescope monitoring from September 1993 and June 2006). 

\begin{figure}[h]
\centering
\includegraphics[width=\columnwidth]{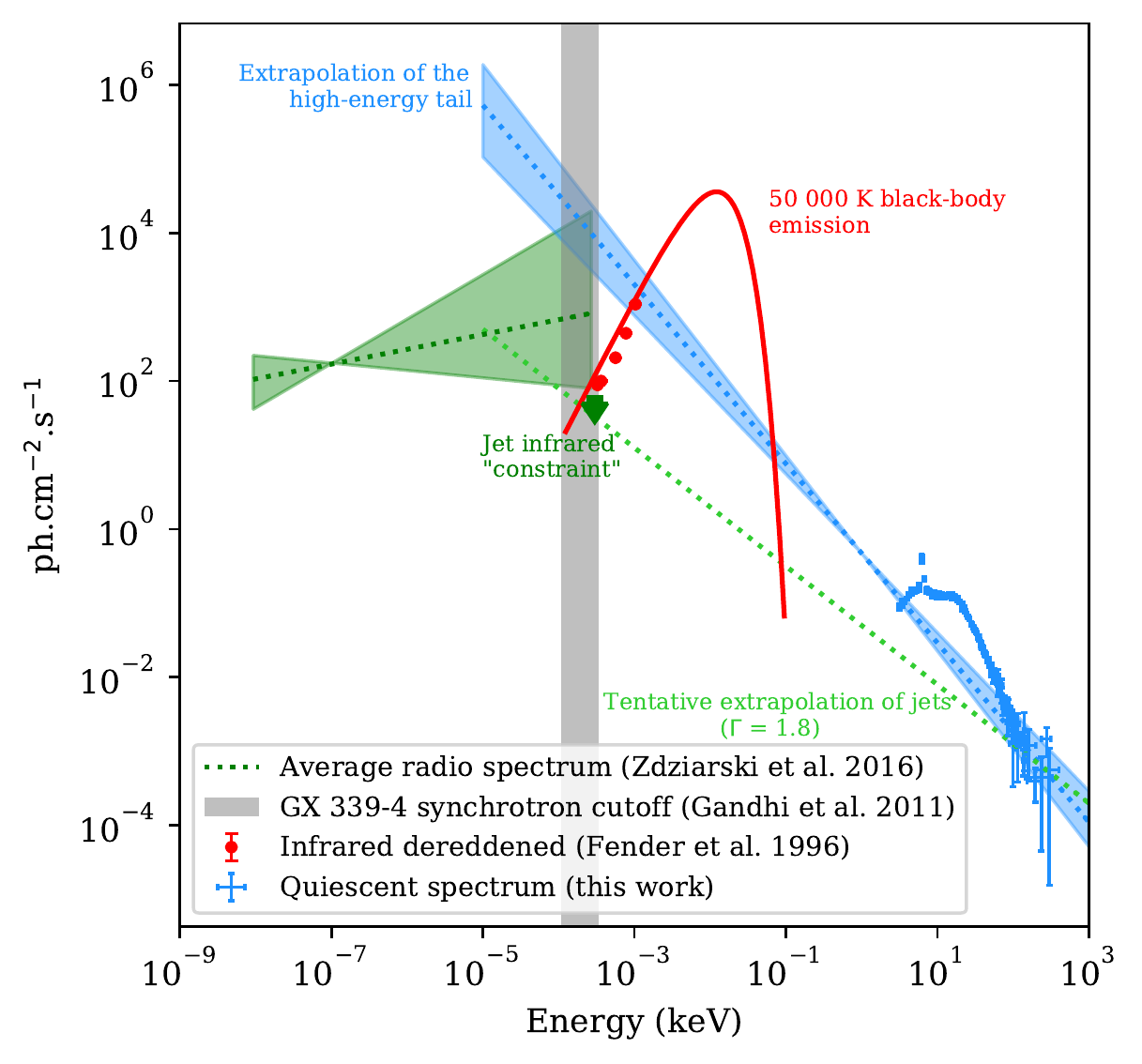}
\caption{Broad-band spectrum of Cyg X-3 in its quiescent state. Radio spectrum, infrared data, and X-ray data are represented with a dark-green dotted line, red dots, and blue dots, respectively. We also show the extrapolation of the high-energy tail to lower energies with a dotted blue line and a black body emission for a temperature of 50 000\,K with a red line. The green arrow shows the rough constraint on the jet synchrotron emission in the infrared
imposed by the detected infrared emission associated with stellar emission \citep{Fender1996}, whereas the light-green dotted line shows the high-energy tail with a photon index of $\Gamma = 1.8$. The gray zone indicates the energy of the synchrotron cut-off for GX 339-4.}
\label{fig:broad_spectrum}
\end{figure}

Figure \ref{fig:broad_spectrum} shows the Cyg X-3 spectral energy distribution (SED) from radio energies to 1000\,keV. We represent a 50 000\,K black-body emission  on the SED (typical temperature of a Wolf-Rayet star) in red. This component is totally consistent with the dereddened infrared points, showing that all measured infrared emission comes from the
companion star \citep{Fender1996}. This allows us to place a rough constraint or limit on the contribution of the jet-synchrotron emission in the infrared (shown with a green arrow), which is necessarily negligible compared to the emission from the star. If one considers the range of the infrared synchrotron break observed in the case of other black hole binaries, for example GX 339-4 \citep[][and constrained between $4.6^{+3.5}_{-2.0} \times 10^{13}$\, Hz in gray in Fig. \ref{fig:broad_spectrum}]{Gandhi2011}, we can extrapolate the then supposed synchrotron emission to the X-rays (green dotted line in Fig. \ref{fig:broad_spectrum}). To reach the high energy, the synchrotron power-law index would need then to be $\Gamma=1.8$, slightly harder than the one we obtain from the spectral fits ($\Gamma = 2.2 \pm 0.1$). Alternatively, extrapolating the high-energy tail down to the infrared domain (blue dotted line in Fig. \ref{fig:broad_spectrum}) results in a much higher infrared flux than measured by \cite{Fender1996}. However, the Fig.\ref{fig:broad_spectrum} shows that the possible synchrotron extension in light green could contribute to the high-energy emission we observe in the X-rays, implying that synchrotron emission could also be a plausible scenario. 

In a recent work, \cite{Pahari2018} used \textit{Astrosat} to measure a rather flat power-law component with a photon index of $1.49^{+0.04}_{-0.03}$ dominating at 20--50\,keV. This component appears during an episode of major ejection and is interpreted as the synchrotron emission from the jets. We do not find such a hard photon index in our FIM state. By doing the same extrapolation of the power law through low energies as these latter authors did, we find a much higher flux (more than ten orders of magnitude) than expected in this state (K10). Nevertheless, the very peculiar event observed by \cite{Pahari2018} may have been smoothed by our approach of stacking spectra. On the other hand, we do observe the hardening of the electron injection index in states where a major ejection is observed, and thus a connection between hard X-rays and radio emission, as previously mentioned.

\subsection{Comparison with previous work}

The global behavior we find with \textsc{eqpair} is similar to that found by H09; the lower the value of $l_\mathrm{h}/l_\mathrm{s}$, the softer the state. Although parameters obtained are globally consistent with the work of H09, we note some differences.

First, we find a different electron injection index in the quiescent/transition ($\Gamma_\mathrm{inj}^\mathrm{Q} = 3.60^{+0.14}_{-0.05}$ and $\Gamma_\mathrm{inj}^\mathrm{T} = 3.31^{+0.14}_{-0.09}$) than in the H09 \textit{hard state }($\Gamma_\mathrm{inj}^\mathrm{hard} = 3.9 \pm 0.1$). Theses differences may come from a better definition at higher energies with \integral, bringing a more precise estimation of $\Gamma_\mathrm{inj}$ than with \textit{RXTE}/HEXTE used by H09.

Secondly and more importantly, our reflection values are weaker than in H09 (particularly in the quiescent and transition states) and we observe a higher value of the ratio $l_\mathrm{nth}/l_\mathrm{h}$ for the quiescent and transition states, leading to a different interpretation of the spectra. Indeed, the bump observed around 20\,keV in the quiescent and transition states is in our case due to Comptonization, and not reflection as in H09. We note that in a previous work on the so-called Cyg X-3 \textit{hard state}, \cite{Hjalmarsdotter2008} came to three slightly different interpretations of their analysis. The results from the present study allow us to break the degeneracy of their interpretation and lead us to favor their nonthermal interpretation \citep[see][for details]{Hjalmarsdotter2008}. In this case, the high value of the ratio $l_\mathrm{nth}/l_\mathrm{h}$ implies that the spectrum is dominated by nonthermal electrons and the peak around 20\,keV is determined by the energy $k\mathrm{T_e}$ at which electrons are injected. This temperature, weaker than observed in other XRBs, is around 4\,keV which means that the peak does not arise from the highest temperature of the electron distribution. Such nonthermal emission could come from shocks due to the dense wind environment, resulting in particle acceleration. Another possibility is that the corona is also the base of the emitting-jet region; in such a geometry, the mechanism responsible for triggering the ejections would also be responsible for the particle acceleration. Whatever the mechanism responsible for this nonthermal emission, it has to be efficient enough in order to prevent the thermal heating of the plasma electrons.

We also note differences with the set of parameters obtained by \cite{Corbel2012} which use \textit{RXTE}/PCA data and the \eqpair model in order to provide some insight into the global evolution of the 3--50\,keV spectrum during a major radio flare. These latter authors in particular find much softer values for the injected electron index and low seed photon temperatures. Nevertheless, the goal of their work is not a detailed spectral analysis, as they obtain several degeneracies within the parameters, and we should not over-interpret these differences. Despite these differences, the global trend of their modeling also shows an increase in $l_\mathrm{h}/l_\mathrm{s}$ and a decrease in $l_\mathrm{nth}/l_\mathrm{h}$ as the source goes from hard to soft states.

\subsection{Dependence on orbital modulation}
\label{sec:orbital_mod}

As Cyg X-3 shows strong orbital modulation, we investigated the potential dependence of the nonthermal emission as a function of the orbital position of the source. In the FIM state, we find a slight difference in the photon index value between inferior and superior conjunctions: $\Gamma^\mathrm{inf} = 2.94 \pm 0.04$ whereas $\Gamma^\mathrm{sup} = 2.72 \pm 0.04$. Previously, \cite{Zdziarski2012} observed this kind of behavior by carrying out a phase-resolved spectral analysis with PCA and HEXTE. Their state number \textquotedblleft\,4\,\textquotedblright\ \cite[from the][classification]{Szostek2008}, which corresponds to our FIM state, is also softer when the source is in superior conjunction. The authors explain this variation by an overly short exposure in this state compared to the others. Here, this argument is no longer valid; our IBIS exposure time in the FIM is highest after the transition state (15380\,s and 13015\,s in inferior and superior conjunction, respectively, compared to 43530\,s and 36150\,s in the transition state). Another interpretation is that this is an effect caused by a higher absorption when Cyg X-3 is behind its companion. With absorption affecting soft X-rays, higher energy photons would not be absorbed and the ratio between the emissions from the two different conjunction phases  would be 1. This would bend the spectrum at low energy, resulting in a harder power law. In order to verify this assumption, we extract the density column value for each state and for each phase bin. However, the uncertainties on this parameter are  too large, preventing us from coming to any firm conclusion.

\subsection{Link with the $\gamma$-ray emission}

At higher energies, in the $\gamma$-ray domain, the extrapolation of the power law in the FIM and FSXR states where $\gamma$-ray emission is detected \citep{Piano2012, Zdziarski2018} leads to weaker flux than detected, and the hard-X-ray emission does not seem directly connected to the $\gamma$ emission. However, this latter has  already been interpreted in the context of a leptonic \citep{Dubus2010, Zdziarski2012, Zdziarski2018} or hadronic scenario \citep{Romero2003, Sahakyan2014}. In the leptonic scenario, this emission comes from Compton scattering of stellar radiation by relativistic electrons from the jets \citep{Cerutti2011, Piano2012, Zdziarski2012, Zdziarski2018}. The hadronic scenario on the other hand predicts $\gamma$-ray emission from the decay of neutral pions produced by proton--proton collisions. In the future, the Cerenkov Telescope Array may bring new constraints on the processes that occur at these energies.

\begin{acknowledgements}
The authors thank the referee for constructive comments that halped to improved the quality of the manuscript. The authors also warmly thank E. Jourdain and J.P Roques for their help on SPIDAI, C.-A. Oxborrow and J. Chenevez for their help with the JEM-X calibration issues, and S. Corbel for useful discussions and suggestions. FC, JR, PL \& RB acknowledge partial funding from the French Space Agency (CNES). FC, JR \& RB acknowledge partial fundings from the French Programme National des Hautes Energies (PNHE). VG is supported through the Margarete von Wrangell fellowship by the ESF and the Ministry of Science, Research and the Arts Baden-W\"urttemberg. Based on observations with \integral, an ESA project with instruments and science data centre funded by ESA member states (especially the PI countries: Denmark, France, Germany, Italy, Switzerland, Spain) and with the participation of Russia and the USA. 
\end{acknowledgements}

\bibliographystyle{aa}
\bibliography{aa_abbrv,mnemonic,bib_CygX-3}
\appendix

\section{PCA and JEM-X simultaneous data}

\begin{table}[h]
\center
\begin{tabular}{l c c c}
\hline
obsID & scw & HR PCA & HR JEM-X \\
\hline
91090-02-01-01 & 0437002300 & 0.260 & 0.256 \\
91090-02-01-07 & 0437002900 & 0.249 & 0.368 \\
91090-02-01-02 & 0437003300 & 0.220 & 0.220 \\
91090-03-01-00 & 0437004200 & 0.148 & 0.169 \\
91090-03-02-00(1) & 0438006700 & 0.107 & 0.135 \\
91090-03-02-00(2) & 0438007000 & 0.120 & 0.124 \\
91090-03-02-00(3) & 0438007300 & 0.112 & 0.123 \\
91090-01-01-00(1) & 0462002500 & 0.191 & 0.163 \\
91090-01-01-00(2) & 0462002700 & 0.215 & 0.221 \\
91090-01-01-00(3) & 0462003000 & 0.256 & 0.265 \\
91090-01-01-01 & 0462003200 & 0.220 & 0.220 \\
94328-01-10-00(1) & 0804009800 & 0.264 & 0.223 \\
94328-01-10-00(2) & 0804009900 & 0.220 & 0.200 \\
95361-01-16-00 & 0989009500 & 0.390 & 0.368 \\
95361-01-17-00 & 0996009000 & 0.647 & 0.687 \\
95361-01-36-00 & 1031005100 & 0.217 & 0.210 \\
\hline
\end{tabular}
\vspace{0.25cm}
\caption{PCA and JEM-X hardness ratio of the 16 simultaneous observations.}
\label{annexe:simult}
\end{table}

\end{document}